\documentstyle[epsf,eqsecnum,aps]{revtex}

\def\3{{}^3\!}
\def\0#1{\stackrel{\scriptscriptstyle (0)}{#1}}
\def\1#1{\stackrel{\scriptscriptstyle (1)}{#1}}
\def\Oll{\frac{\nabla_l\nabla^l\Omega}{\Omega}}
\def\Oij{\frac{\nabla_i\nabla_j\Omega}{\Omega}}
\def\OlOl{\frac{\nabla_l\Omega\nabla^l\Omega}{\Omega^2}}
\def\OiOj{\frac{\nabla_i\Omega\nabla_j\Omega}{\Omega^2}}
\def\a{a}
\def\eff{{\mbox{\it eff}}}
\makeatletter
\def\figurenum#1{\def\thefigure{#1}\let\@currentlabel\c@thefigure
\addtocounter{figure}{\m@ne}}
\def\fnum@figure{{\rm Fig.\space\thefigure.}}
\makeatother
\begin{document}

\begin{flushright}
TIT/HEP-326/COSMO-71 \\
\today
\end{flushright}
\begin{center}
{\bf
\vskip 0.25cm
{\Large Onset of inflation in inhomogeneous cosmology}}\\
\vskip 0.8cm

Osamu Iguchi
\footnote{e-mail: osamu@th.phys.titech.ac.jp, \quad JSPS junior fellow}
and Hideki Ishihara
\footnote{e-mail: ishihara@th.phys.titech.ac.jp}\\
\vskip 0.8cm
{ Department of Physics \\
  Tokyo Institute of Technology, Meguroku, Tokyo 152, Japan}
\vskip 0.5cm

{\bf abstract}
\vskip 0.15cm

\begin{minipage}[c]{9.5cm}
\small

We study how the initial inhomogeneities of the universe 
affect the onset of inflation in the closed universe. 
We consider the model of a chaotic inflation which is driven by
a massive scalar field. 
In order to construct an inhomogeneous universe model, 
we use the long wavelength approximation 
( the gradient expansion method ). 
We show the condition of the inhomogeneities for the universe to
enter the inflationary phase. 

PACS number(s): 98.80.Cq, 98.80.Hw
\end{minipage}
\end{center}
\vspace{0.5cm }

\section{Introduction}
\label{sec:intro}

The main objective of inflationary scenario\cite{Inf} 
is to explain why
the present universe is homogeneous and isotropic on large
scales without a fine tuning of initial conditions.
It is important to study the generality of 
initial conditions for the onset of inflation.

Many works on inflation assume homogeneous and flat universe
at the pre-inflationary phase\cite{GP1}.
In this case, inflation is generic.
The spatial curvature effect on the occurrence of inflation
is investigated in the Friedmann-Robertoson-Walker 
model with a massive scalar field by Belinsky {\it{et al}}\cite{BIK},
and in the Bianchi IX model with a cosmological constant by Wald\cite{Wald}.
They concluded that inflation is not a general property and
only the positive spatial curvature
can prevent the universe from inflating.

On the other hand, the role of initial inhomogeneities on
the occurrence of inflation was studied by a linear perturbation
analysis \cite{FB} and by a numerical simulation
with special symmetries\cite{GP2,SM}.
In ref. \cite{GP2}, Goldwirth and Piran concluded that the crucial
feature necessary
for inflation is a sufficient high average value of the scalar field
which drives inflation over a region of
several Hubble radius.

We study how the initial inhomogeneities of the universe
affect the onset of inflation by use of an alternative approach.
In order to treat the inhomogeneities of the universe,
we use the long wavelength approximation.

In the long wavelength approximation 
we take a synchronous reference frame
where the line element is of the form
\begin{eqnarray}
 ds^2 = g_{\mu\nu}dx^{\mu}dx^{\nu}
      = -dt^2+\gamma_{ij}(t,x) dx^idx^j .
\label{eq:synch}
\end{eqnarray}
Throughout this paper Latin letters will denote spatial indices and
Greek letters spacetime indices. 
In the usual method, 
we neglect all spatial gradients of $\gamma_{ij}(t,x)$
in the Einstein equations and construct an approximate solution with
the characteristic scale of inhomogeneities larger than 
the Hubble radius\cite{LK,Tomita,Comer,Salopek,DL}.
Furthermore, we consider the quasi-isotropic universe 
as a special case in the form
\begin{eqnarray}
 ds^2 = -dt^2+a^2(t)h_{ij}(x)dx^idx^j , 
\label{eq:qasiiso}
\end{eqnarray}
where $h_{ij}(x)$ is an arbitrary function 
called \lq seed metric\rq . 
The universe is assumed to be filled with a perfect fluid 
characterized by energy density $\rho$
and pressure $p=(\Gamma-1) \rho$. 
In this case the Einstein equations reduce to
\begin{eqnarray}
        &&\frac{\ddot a}{a}
                -\frac{2-3\Gamma}{2}\left(\frac{\dot a}{a}\right)^2
                = 0, 
\label{eq:F-eq1} \\
        &&\left(\frac{\dot a}{a}\right)^2
         = \frac{1}{3}\kappa\rho ,
\label{eq:F-eq2}
\end{eqnarray}
where a dot denotes the derivative with
respect to $t$ and $\kappa\equiv 8\pi G$.
These are nothing but the equations which the scale factor $a(t)$
of the flat Friedmann universe should satisfy.
We call the approximate inhomogeneous solution given by Eq.(\ref{eq:qasiiso})
satisfying Eqs.(\ref{eq:F-eq1}) and (\ref{eq:F-eq2})
the locally flat Friedmann solution and it 
is used as a starting point to solve the Einstein 
equations iteratively\cite{Comer,Salopek}.
In the iteration scheme, the spatial metric can be expanded
as a sum of spatial tensors of increasing order in spatial
gradients of $h_{ij}(x)$.
Thus the approximation is called the gradient expansion method.

Recently, the influence of initial inhomogeneities
on the occurrence of
inflation is studied by using the gradient expansion method
\cite{NG,Numbu,IIS}.
From investigations in the homogeneous universe model,
as noted above, the effect of positive curvature is
important because it can prevent inflation. 
So, we should investigate the inhomogeneities of universe
with a non-small positive spatial curvature to clarify 
the generality of inflation. 
However, the lowest order solution in the gradient expansion method
is a locally flat Friedmann universe and the spatial curvature is
treated as a small quantity in the expansion scheme.

In ref.\cite{IIS}, 
the gradient expansion method from 
the locally closed Friedmann universe is introduced 
and the inhomogeneous closed universe with a 
cosmological constant 
is studied as an application of the scheme. 
This method is useful to treat the inhomogeneities of non-small 
curvature associated with a three-space which is 
conformal to the constant curvature space at lowest order. 
In this paper we consider inflation which is driven by 
a massive scalar field 
in the inhomogeneous universe by use of 
the first order approximation of the gradient expansion from 
the locally closed Friedmann universe. 
We will show the condition of the inhomogeneities of the spatial
curvature for the universe to enter the inflationary phase.

This paper is organized as follows.
In Sec.2, we derive basic equations 
by use of the first order approximation in the gradient
expansion method from the locally closed Friedmann universe.
In Sec.3, we show numerical results.
Sec.4 is devoted to conclusion.
We use units in which $c =\hbar=1$.

\section{Basic equations in improved gradient expansion method}
\label{sec:beq}

We take the matter in the universe to be a scalar field minimally
coupled to gravity.
The energy-momentum tensor of a scalar field is given by
\begin{eqnarray}
	T_{\mu\nu} &=& \phi_{,\mu}\phi_{,\nu}-g_{\mu\nu}\left[
                \frac{1}{2}\phi_{,\lambda}\phi^{,\lambda}+V(\phi)
                \right],
\end{eqnarray}
where $V(\phi)$ is the potential of the scalar field.
In the synchronous reference frame Eq.(\ref{eq:synch}),
the Einstein equations for
the gravitational field coupled to the scalar field read
\begin{eqnarray}
 \frac{1}{2}\dot{K}+\frac{1}{4}K^l_mK^m_l &=&
        \kappa\left[ -\dot\phi^2+V(\phi)\right], \label{eq:ein1}\\
 \3R^j_i+\frac{1}{2\sqrt{\gamma}}
 \frac{\partial}{\partial t}(\sqrt{\gamma} K^j_i) &=&
        \kappa \left[\partial_i\phi\partial^j\phi+V(\phi)\delta^j_i\right],
 \label{eq:ein2}\\
 \frac{1}{2}(K^j_{i;j}-K_{,i}) &=& \kappa\dot\phi\partial_i\phi,
        \label{eq:ein3}
\end{eqnarray}
where $\3R_{i}^{j}$ is the Ricci tensor associated with $\gamma_{ij}$,
$K_{ij}\equiv\dot\gamma_{ij}$, $K\equiv\gamma^{ij}K_{ij}$,
$\gamma\equiv\det\gamma_{ij}$,
and a semicolon denotes the covariant derivative with
respect to $\gamma_{ij}$.
The equation of motion for the scalar field is
\begin{eqnarray}
 \ddot\phi+\frac{1}{2}K\dot\phi-\phi^{;l}_{;l}+\frac{d V(\phi)}{d\phi}&=& 0.
 \label{eq:seom}
\end{eqnarray}

Now, we assume the metric in the form\cite{IIS} 
\begin{eqnarray}
 \0\gamma_{ij}(t,x) &=& a^2(t,\Omega(x))h_{ij}(x),
 \label{eq:metric0}
\end{eqnarray}
where $h_{ij}(x)$ is the metric of three-dimensional sphere
whose Ricci curvature is $R_{ij}(h_{kl})=2h_{ij}$.  
The scale factor $a(t,\Omega(x))$ in the spatial metric Eq.(\ref{eq:metric0})
has the space-dependence by an arbitrary function $\Omega(x)$.
In the usual gradient expansion method,
all spatial gradients of $\0\gamma_{ij}(t,x)$
are assumed to be much smaller than time derivatives.
Roughly speaking, the spatial curvature $\3R$ is assumed to be smaller than
$K^2$ and it is neglected in the lowest approximation.
Here, we consider slightly different approximation.
The spatial scalar curvature for the metric in the form of
Eq.(\ref{eq:metric0}) is
\begin{eqnarray}
 \3R(\0\gamma_{ij}) &=& \frac{6}{a^2(t,\Omega)}
        +\frac{1}{a^2(t,\Omega)}\left[
        -4\frac{a'}{a}\Omega\Oll 
        +2\left(\frac{a'^2}{a^2}-2\frac{a''}{a}\right)\Omega^2
                \OlOl\right] ,
\label{eq:3R}
\end{eqnarray}
where a prime denotes the derivative with respect to $\Omega$
and $\nabla_l$ denotes the covariant derivative with respect to $h_{ij}$. 
We assume that the second term in the right hand side of Eq.(\ref{eq:3R})
which contains spatial gradients of $\Omega$ is smaller than
the first term.
If we set $\3{\0R}(\0\gamma_{ij}) \equiv 6 a^{-2}(t,\Omega)$
Eq.(\ref{eq:3R}) is rewritten in the form
\begin{equation}
        \3R(\0\gamma_{ij}) = \3{\0R}(\0\gamma_{ij})
                +\left[
                   2\frac{\3{\0R}{}_{;l}^{~;l}}{\3{\0R}}
                  -\frac{3}{2} \frac{\3{\0R}_{;l}~\3{\0R}{}^{;l}}{\3{\0R}{}^2}
                \right] .
\label{eq:3R2}
\end{equation}
Thus the case, we consider here, is that the spatial variation of 
the spatial curvature is smaller than the value of itself. 
All spatial gradients of $\Omega(x)$ is neglected in the first
step of our approach.
It should be noted that even if the second term 
in Eq.(\ref{eq:3R}) or (\ref{eq:3R2}) exceeds 
the first term, the approximation reduces to the usual long wavelength 
approximation so far as these are smaller than $K^2$.

Though the metric in the form of Eq.(\ref{eq:metric0}) has not adequate
degree of gravitational freedom, the simple form 
on which we concentrate is useful to investigate 
the inhomogeneities of the spatial curvature. 
It might be very complicate to treat the general form of inhomogeneities 
of curvature\cite{DL}. 

From now on, we consider that the spatial metric
in the lowest order are described by Eq.(\ref{eq:metric0}). 
Substituting Eq.(\ref{eq:metric0}) into Eq.(\ref{eq:ein1}), 
and neglecting all spatial derivatives of $\Omega(x)$, 
we obtain the equation which $a$ and $\phi$ must satisfy: 
\begin{eqnarray}
 \frac{\ddot{a}(t, \Omega(x))}{a(t, \Omega(x))}
        &=& \frac{\kappa}{3}\left[-\dot\phi^2(t,x)+V(\phi(t,x))\right] . 
\label{eq:sa0m}
\end{eqnarray}
From Eq.(\ref{eq:sa0m}), it is natural that the scalar field has
the space-dependence through $\Omega(x)$.
We assume that the form of the scalar field in the lowest order is
\begin{eqnarray}
        \0\phi\!(t,x) = \phi_0(t,\Omega(x)).
\label{eq:sscalar0}
\end{eqnarray}
Substituting Eqs.(\ref{eq:metric0}) and (\ref{eq:sscalar0})
into Eqs.(\ref{eq:ein2}) and (\ref{eq:seom}), 
and neglecting all spatial derivatives of $\Omega(x)$,
we have
\begin{eqnarray}
 \frac{\ddot a(t,\Omega)}{a(t,\Omega)}
 +2\left(\frac{\dot a(t,\Omega)}{a(t,\Omega)}\right)^2
 +\frac{2}{a^2(t,\Omega)} &=& \kappa V(\phi_0(t,\Omega)),
 \label{eq:sa0c}\\
 \ddot\phi_0(t,\Omega)
 +3\frac{\dot{a}(t,\Omega)}{a(t,\Omega)}\dot\phi_0(t,\Omega)
 +\frac{d V(\phi_0(t,\Omega))}{d\phi_0(t,\Omega)} &=& 0.
 \label{eq:sphi0}
\end{eqnarray}
Equation(\ref{eq:ein3}) is trivial in the lowest order. 
Equations(\ref{eq:sa0m}), (\ref{eq:sa0c}) and (\ref{eq:sphi0}) 
have the same form of the equations for the closed
Friedmann universe with a homogeneous scalar field.

At next order, we consider the second order in spatial
gradients of $\Omega(x)$.
We will take corrections to the metric and the scalar field of the form
\begin{eqnarray}
 \1{\gamma_{ij}}(t,x) 
	&=& a^2(t,\Omega)\left[ 
		\frac{1}{3}F(t,\Omega)\Oll h_{ij}
 		+\overline F(t,\Omega)\overline{\Oij} \right.\nonumber\\
 		&&\qquad\quad \left.+\frac{1}{3}G(t,\Omega) \OlOl h_{ij}
 		+\overline G(t,\Omega)\overline{\OiOj}\right], 
\label{eq:metric1}\\
 \1\phi(t,x) &=&  P(t,\Omega)\Oll +Q(t,\Omega)\OlOl ,
\label{eq:sscalar1}
\end{eqnarray}
where $\overline{\nabla_{i}\nabla_{j}\Omega/\Omega}$ and
$\overline{\nabla_{i}\Omega\nabla_{j}\Omega/\Omega^2}$ are traceless
tensors defined by
\begin{eqnarray*}
 \overline{\Oij} \equiv \Oij - \frac{1}{3}\Oll h_{ij}, \quad 
 \overline{\OiOj} \equiv \OiOj -\frac{1}{3}\OlOl h_{ij}.
\end{eqnarray*}

Substituting $\gamma_{ij}=\0{\gamma}_{ij}+\1{\gamma}_{ij}$ 
and $\phi = \0\phi + \1\phi$
given by Eqs.(\ref{eq:metric0}), (\ref{eq:metric1}), 
(\ref{eq:sscalar0}) and (\ref{eq:sscalar1}) 
into the Einstein equations 
and the equation of motion for the scalar field
Eqs.(\ref{eq:ein1})--(\ref{eq:seom}), 
and comparing the coefficients of the derivative of $\Omega$,
we get the first order equations which govern the evolution 
of variables $F, \overline{F}, G, \overline{G}, P$ and $Q$. 
We should perform tedious calculation to get $\3R^i_j$ associated with 
$\gamma_{ij}=\0{\gamma}_{ij}+\1{\gamma}_{ij}$. 
The expression of it in the first order approximation 
appears in Appendix \ref{sec:rij}. 

%
\def\V'{\frac{dV}{d\phi}\Big\vert_0}
From Eq.(\ref{eq:ein2}) we obtain
\begin{eqnarray}
 \ddot{F}+6H\dot{F} &=& 6\kappa \V' P
        +\frac{4}{a^2}\left[F-\overline{F}+2\frac{a'}{a}\Omega\right],
        \label{eq:shcf} \\
 \ddot{\overline{F}}+3H\dot{\overline{F}} &=& \frac{2a'}{a^3}\Omega,
        \label{eq:shcf-} \\
 \ddot{\tilde{G}}+6H\dot{\tilde{G}} &=&
        6\kappa \V' Q+\frac{2\kappa}{a^2}{\phi'}_0^2\Omega^2
        +\frac{4}{a^2}\left[\tilde{G}+\overline{F}-\overline{F}'\Omega
        +\left(2\frac{a''}{a}-\frac{a'^2}{a^2}\right)\Omega^2\right],
        \label{eq:shcg}\\
 \ddot{\hat{G}}+3H\dot{\hat{G}} &=&
        \frac{2\kappa}{a^2}\phi'^2_0\Omega^2
        +\frac{2}{a^2}\left[2\hat{G}+2\overline{F}
        -2\overline{F}'\Omega-\frac{3}{2}\overline{F}^2
        +\left(\frac{a''}{a}-\frac{2a'^2}{a^2}
        \right)\Omega^2\right], \label{eq:shcg-}
\end{eqnarray}
where $\displaystyle\V'$ is the lowest order value of 
$\displaystyle\frac{dV}{d\phi}$, 
$H\equiv \dot{a}/a$, $\tilde{G}\equiv G+\overline F^2$ and 
$\hat G \equiv \overline G -\frac{1}{2}{\overline F}^2$.
%
Note readers should pay attention to the commutation relation of
the covariant derivatives which is not vanishing because the spatial
surface in the lowest order solution has a positive curvature.
From Eq.(\ref{eq:seom}), we have
\begin{eqnarray}
 \ddot P +3H\dot P+\frac{d^2V}{d\phi^2}\Big\vert_0 P 
	&=& -\frac{1}{2}\dot\phi_0\dot{F}
        +\frac{1}{a^2}\phi'_0\Omega, 
\label{eq:ssmp} \\
 \ddot Q+3H\dot Q+\frac{d^2V}{d\phi^2}\Big\vert_0 Q 
	&=& -\frac{1}{2}\dot\phi_0\dot{\tilde{G}}
        +\frac{1}{a^2}\left[\phi''_0+\frac{a'}{a}\phi'_0\right]\Omega^2.
\label{eq:ssmq}
\end{eqnarray}
From Eqs.(\ref{eq:ein1}) and (\ref{eq:ein3})  we obtain
\begin{eqnarray}
 \dot{\overline{F}} &=&
        2\left[\frac{\dot a'}{a}-\frac{\dot{a}a'}{a^2}\right]\Omega
        +\kappa \dot\phi_0\phi'_0\Omega ,
\label{eq:smc1} \\
 \ddot{F}+2H\dot{F} &=&
        2\kappa\left[-2\dot\phi_0\dot P+\V'P\right],
 \label{eq:smf}  \\
 \ddot{\tilde{G}}+2H\dot{\tilde{G}} &=&
        2\kappa\left[-2\dot\phi_0\dot Q+\V' Q\right]
        +\dot{\overline{F}}^2 .
\label{eq:smg}
\end{eqnarray}

There are nine equations (\ref{eq:shcf})--(\ref{eq:smg})
for the six variables $F, \overline{F}, \tilde{G}, \hat{G}, P$ and $Q$.
All of these equations are not independent. 
Six equations govern the evolution and three are constraint conditions. 
So we find a solution of Eqs.(\ref{eq:shcf})--(\ref{eq:ssmq}) with 
an initial condition which satisfies the constraints. 
In order to solve these equations, 
we need to know the time evolution of $a', a'', \phi_0', \phi_0''$
and $\overline{F}'$. 
Differentiating Eqs.(\ref{eq:sa0m}), (\ref{eq:sa0c}), (\ref{eq:sphi0})
and (\ref{eq:smc1}), we obtain the equations for them, 
Eqs.(\ref{eq:a'})--(\ref{eq:overF'}), which appear in 
Appendix \ref{sec:subeq}. 
At higher order, we consider the corrections which are constructed by
the spatial derivatives of $\Omega(x)$ and solve the Einstein equations
order by order in the gradient expansion.
The spatial metric and the scalar field can be expanded as
\begin{eqnarray}
        \gamma_{ij}(t,x)&=& a^2(t,\Omega)\left[ h_{ij}
                + \sum_A F_{(2)}^A(t,\Omega)(\nabla^{(2)}\Omega)^A_{ij}
                + \sum_A F_{(4)}^A(t,\Omega)(\nabla^{(4)}\Omega)^A_{ij}
                + \cdots \right. \nonumber \\
                &&\left. \qquad \qquad \quad ~~
                + \sum_A F_{(2p)}^A(t,\Omega)(\nabla^{(2p)}\Omega)^A_{ij}
                + \cdots \right] ,
        \label{eq:exp_metric} \\
        \phi(t,x)&=& \phi_0(t,\Omega)
                + \sum_A P_{(2)}^A(t,\Omega)(\nabla^{(2)}\Omega)^A_{ij}
                + \sum_A P_{(4)}^A(t,\Omega)(\nabla^{(4)}\Omega)^A_{ij}
                + \cdots \nonumber \\
                && \qquad \quad ~~ 
                + \sum_A P_{(2p)}^A(t,\Omega)(\nabla^{(2p)}\Omega)^A_{ij}
                + \cdots ,
        \label{eq:exp_phi}
\end{eqnarray}
where the notation $(\nabla^{(2p)}\Omega)^A_{ij}$ denotes symbolically
symmetric spatial tensors which contain $2p$ spatial gradients
of $\Omega(x)$, the suffix $A$ distinguishes the tensors belonging
to the same class, and $a(t,\Omega)$, $\phi_0(t,\Omega)$,
$F_{(2p)}^A(t,\Omega)$ and $P_{(2p)}^A(t,\Omega)$ are function of $t$
and $\Omega(x)$ which is determined by the Einstein equations.
In the case of $p=1,2$, the explicit form of $(\nabla^{(2p)}\Omega)^A_{ij}$
appears in the Appendix \ref{sec:tensor}.

This expansion scheme is valid when the characteristic scale of
inhomogeneities of the spatial curvature is larger than the
Hubble radius $(\dot a/a)^{-1}$ or the spatial curvature radius
$\3\!\0R\!{}^{-1/2}$.
If the Hubble radius is smaller than the spatial curvature radius,
this expansion scheme reduces to the usual gradient expansion.

\section{The influence of inhomogeneities for inflation}
\label{sec:scalar}

We consider a chaotic inflation model with a massive scalar field,
{\it i.e.,} $V(\phi)=\frac{1}{2}m^2\phi^2$
as an inflaton field.

When the energy of the scalar field is dominated by a kinetic term
at the early stage,
the initial behavior of solution without inhomogeneities
for Eqs.(\ref{eq:sa0m}),
(\ref{eq:sa0c}) and (\ref{eq:sphi0}) is
expressed by series expansions in the form
\begin{eqnarray}
 a^2(t) &=& m^{-2} t^{\frac{2}{3}}\left[1
        -\frac{9}{7}t^{\frac{4}{3}} + O(t^\frac{8}{3})\right],
\label{eq:a-homo} \\
 \phi_0(t) &=& \sqrt{\frac{2}{3\kappa}}
        \left[\ln t +\frac{81}{56}t^{\frac{4}{3}} + O(t^\frac{8}{3})
        \right] +\tilde{\phi}_0,
\label{eq:phi-homo}
\end{eqnarray}
where  $\tilde{\phi}_0$ is the constant
which corresponds to the freedom of initial value of
the scalar field. 
The time variable $t$ is non-dimensionalized by use of $m$.
On assumption that we neglect all spatial derivatives of $\Omega$, 
we can generalize 
the solution Eqs.(\ref{eq:a-homo}) and (\ref{eq:phi-homo}) in the form
\begin{eqnarray}
 a^2(t,\Omega(x))
        &=& \Omega^2(x) t^{\frac{2}{3}} \left[1
        -\frac{9m^{-2}}{7\Omega^2(x)}t^{\frac{4}{3}}
        + O(\frac{m^{-4}}{\Omega^4}t^\frac{8}{3})  \right], \\
 \phi_0(t,\Omega(x)) &=& \sqrt{\frac{2}{3\kappa}}
        \left[\ln t +\frac{81m^{-2}}{56\Omega^2(x)}t^{\frac{4}{3}}
        + O(\frac{m^{-4}}{\Omega^4}t^\frac{8}{3}) \right]
        +\tilde{\phi}_0 . 
\end{eqnarray}
The local scale factor and the scalar field
are expressed by series expansions of $(m^{-2}/\Omega^2)t^\frac{4}{3}$ 
in the initial stage. 
At the beginning of the universe, the local scale factor behaves as 
$a(t,\Omega) \propto t^{1/3}$ and after that the time evolution depends 
on the local value of $\Omega(x)$. 
In the initial stage where the kinetic energy of scalar field 
dominates the potential energy, 
the behavior of $a(t,\Omega)$ does not depend on $\tilde{\phi}_0$ 
but in the late stage where the potential energy becomes effective, 
it depends on $\tilde{\phi}_0$. 

We can see the time evolutions of these variables, 
which is characterized by
the value of 
$\Omega(x)$ and $\tilde{\phi}_0$, 
by solving Eqs.(\ref{eq:sa0m}), (\ref{eq:sa0c}) 
and (\ref{eq:sphi0}) numerically. 
Fig.\ref{fig:cs0} shows the inflationary region and the recollapsing 
one on $(\Omega^2, \tilde{\phi}_0)$ plane. 
The local scale factor with parameters in the inflationary region enters
the accelerating expansion phase:
$\dot{a}/a>0$ and $\ddot{a}/a>0$.
On the other hand, the local scale factor with parameters in the recollapsing
region enters the recollapsing phase:
$\dot{a}/a<0$ and $\ddot{a}/a<0$.
From Fig.\ref{fig:cs0}, we see that
inflation favors large $\Omega$ and large $|\tilde{\phi}_0|$ 
as is expected by the investigation of the homogeneous model\cite{BIK}.
For fixed $\tilde{\phi}_0$ there is a critical value of $\Omega$,
$\Omega_{cr}$,
if $\Omega>\Omega_{cr}$ inflation occurs. 
Since the critical value $\Omega_{cr}$ depends on $\tilde{\phi}_0$ then
the condition for the onset of inflation in the lowest order solution is
written in the form 
\begin{eqnarray}
 \frac{1}{\Omega^2}<\frac{1}{\Omega_{cr}^2(\tilde{\phi}_0)}.
 \label{criterion01}
\end{eqnarray}
In the Fig.\ref{fig:cs0} $\Omega_{cr}(\tilde{\phi}_0)$ is drawn as 
the boundaries between the inflationary region 
and the recollapsing region.

In the initial stage where the local scale factor $a(t,\Omega(x))$ is
proportional to $\Omega(x)t^{\frac{1}{3}}$, the spatial
scalar curvature behaves as
$\Omega^{-2}t^{-\frac{2}{3}}$.
The condition (\ref{criterion01}) is, then, translated to 
the condition on the initial curvature as
\begin{eqnarray}
        \3R_{init}(t, \Omega) < \3R_{cr}(t, \tilde\phi_0),
\label{criterion02}
\end{eqnarray}
where $\3R_{cr}\equiv 6t^{-\frac{2}{3}}\Omega^{-2}_{cr}(\tilde\phi_0)$.
This condition for the onset of inflation is a restriction on only
the local value of $\3R$. 

Next, we take the first order correction terms into account.
We integrate Eqs.(\ref{eq:shcf})--(\ref{eq:ssmq}) and 
Eqs.(\ref{eq:a'})--(\ref{eq:overF'}) numerically.
We restrict ourselves to consider growing solution whose 
initial asymptotic behavior is
\begin{eqnarray*}
 F=\frac{18 m^{-2}}{7\Omega^2}t^{\frac{4}{3}},
 \bar F=\frac{9 m^{-2}}{8\Omega^2}t^{\frac{4}{3}},
 \tilde{G}=-\frac{9m^{-2}}{7\Omega^2}t^{\frac{4}{3}},
 \hat{G}=-\frac{9m^{-2}}{4\Omega^2}t^{\frac{4}{3}},
 P=-\sqrt{\frac{3}{2\kappa}}\frac{9m^{-2}}{14\Omega^2}t^{\frac{4}{3}},
 Q=\sqrt{\frac{3}{2\kappa}}\frac{9m^{-2}}{28\Omega^2}t^{\frac{4}{3}}.
\end{eqnarray*}
The time evolution of $F$, $\tilde{G}$, $P$ and $Q$ is shown 
in Figs.\ref{fig:F}--\ref{fig:Q}.

In the case that the local scale factor in the lowest order 
will enter the inflationary phase ($\Omega^2 > \Omega^2_{cr}$), 
$F$ and $\tilde{G}$ grow in proportion to time, and 
$P$ and $Q$ approach constant values.
On the other hand, in the case that the local scale factor in the lowest order
will enter the recollapsing phase ($\Omega^2 < \Omega^2_{cr}$),
$F$, $\tilde{G}$, $P$ and $Q$ diverge and the approximation
breaks down in the course of time.
These behavior are consistent with the results by the linear perturbation
analysis \cite{FB}.
The trajectory of scalar field in  $(\phi,\dot\phi)$ space
is shown in Figs.\ref{fig:phia}--\ref{fig:phib}.

At each spatial point, we define an effective local scale factor $a_\eff(t,x)$ 
by
\begin{eqnarray}
 a_\eff(t,x)
        &\equiv& \left[\det\left[\0\gamma_{ij}(t,x)
                +\1\gamma_{ij}(t,x)\right]\right]^{1/6} \nonumber\\
        &=& a(t,\Omega)\left[ 1 +\frac{1}{6}F(t,\Omega)\Oll
                +\frac{1}{6}\tilde{G}(t,\Omega)\OlOl \right] ,
\end{eqnarray}
which describes how the universe expands at the spatial point. 
We can follow the evolution of $a_\eff(t,x)$ by 
the evolution of $a(t,\Omega)$, $F(t,\Omega)$ and $\tilde{G}(t,\Omega)$. 
The behavior of $a_\eff(t,x)$ is characterized by four parameters
$\tilde{\phi}_0, \Omega, \nabla_l\nabla^l\Omega/\Omega$
and $\nabla_l\Omega\nabla^l\Omega/\Omega^2$.
In addition to the value of $\tilde{\phi}_0$ and $\Omega$,  
the spatial derivatives of $\Omega$ affect the evolution of the 
local scale factor. 

The effective local expansion rate and acceleration rate are given by
\begin{eqnarray*}
 \frac{\dot{a}_\eff(t,x)}{a_\eff(t,x)}
        &=& \frac{\dot a(t,\Omega)}{a(t,\Omega)}
                +\frac{1}{6}\left[ \dot F(t,\Omega)\Oll
                +\dot{\tilde{G}}(t,\Omega)\OlOl \right] ,\\
 \frac{\ddot{a}_\eff(t,x)}{a_\eff(t,x)}
        &=& \frac{\ddot{a}(t,\Omega)}{a(t,\Omega)}
                 +\frac{1}{6}\left[\ddot F(t,\Omega)
              +2\frac{\dot{a}(t,\Omega)}{a(t,\Omega)}\dot{F}(t,\Omega)
                 \right]\Oll
              +\frac{1}{6}\left[\ddot{\tilde{G}}(t,\Omega)
              +2\frac{\dot{a}(t,\Omega)}{a(t,\Omega)}\dot{\tilde{G}}(t,\Omega)
                 \right]\OlOl . 
\end{eqnarray*}
We can divide the four-dimensional parameter space 
$(\tilde{\phi}_0, \Omega, 
\nabla_l\nabla^l\Omega/\Omega, \nabla_l\Omega\nabla^l\Omega/\Omega^2)$ 
into two regions: 
inflationary region and recollapsing region by a three dimensional 
hyper surface. 
The effective local scale factor with the parameters in the inflationary 
region enters the accelerating expansion phase: 
$\dot{a}_\eff/a_\eff>0$ and $\ddot{a}_\eff/a_\eff>0$.
On the other hand, the effective local scale factor with the parameters in the
recollapsing region enters the recollapsing phase: 
$\dot{a}_\eff/a_\eff<0$ and $\ddot{a}_\eff/a_\eff <0$.
We assume the approximation is valid while 
\begin{eqnarray}
 \left|\frac{\det\left(\0\gamma_{ij}+\1\gamma_{ij}\right)-\det\0\gamma_{ij}}
        {\det\0\gamma_{ij}}\right| < 0.5 \quad \mbox{and}\quad
 \left|\frac{\3\!\1R(\gamma_{ij})}
        {\max\left\{ H^2, \3\!\0R(\gamma_{ij})\right\}}\right|
        < 0.5 .
\label{eq:ce}
\end{eqnarray}
We evolve the universe numerically 
while both of these conditions Eq.(\ref{eq:ce}) hold.

Evolving of the effective local expansion rate 
and acceleration rate numerically, we see how 
the two spatial gradients, 
$\nabla_l\nabla^l\Omega/\Omega$ and $\nabla_l\Omega\nabla^l\Omega/\Omega^2$, 
affect the occurrence of inflation. 
For fixed $\tilde{\phi}_0$, the inflationary region and recollapsing 
region are  shown in Figs.\ref{fig:cs2502}--\ref{fig:cs2598}. 
Between these two regions, there is a fuzzy region 
where error becomes large and one of inequality in Eq.(\ref{eq:ce}) 
does not hold before the effective local scale factor enters 
the accelerating expansion phase or the recollapsing phase. 
We see that the positive (negative) $\nabla_l\nabla^l\Omega/\Omega$ helps 
the universe to enter the inflationary (recollapsing) phase. 
On the other hand, $\nabla_l\Omega\nabla^l\Omega/\Omega^2$ 
tends to prevent the onset of inflation.

From the numerical results, we obtain the time independent 
condition for the onset of 
inflation in the vicinity of $\Omega/\Omega_{cr}-1=0$ and 
$\nabla_l\nabla^l\Omega/\Omega=\nabla_l\Omega\nabla^l\Omega/\Omega^2=0$ 
in the parameter space, 
within the first order of gradient expansion, 
\begin{equation}
 \frac{1}{\Omega^2}\left[1 +\alpha\left(-\Oll+\beta\OlOl \right)\right]
 	< \frac{1}{\Omega^2_{cr}(\tilde\phi_0)},
\label{criterion1}
\end{equation}
where $\alpha$ and $\beta$ are constants.
The parameter $\alpha$ shows the strength of the effect of 
inhomogeneities on the occurrence 
of inflation. 
In the massive scalar inflaton model, the value of $\alpha$ is about 2.0.
The value of $\beta$ depends on the initial value of 
the scalar field $\tilde{\phi}_0$: 
$\beta=4.3$ for $\tilde\phi_0=5.0\kappa^{-1/2}$ 
and $\beta=3.3$ for $\tilde\phi_0=10.0\kappa^{-1/2}$. 
In the early stage where $a^2(t,\Omega)=\Omega^2t^{\frac{2}{3}}$, 
the spatial curvature $\3R_{init}$ in the 
first order approximation is
\begin{equation}
 \3R_{init}(t, \Omega)
	=\frac{2}{t^{\frac{2}{3}}\Omega^2}\left[ 3 - 2 \Oll +\OlOl\right]. 
\label{Rinit}
\end{equation}
By use of Eq.(\ref{Rinit}), the criterion (\ref{criterion1}) 
is rewritten as 
\begin{equation}
 \3R_{init}(t, \Omega)
 +\tilde\alpha\left\{ \frac{(\3R_{init})_{;l}^{~;l}}{\3R_{init}}
 +\tilde\beta  \frac{(\3R_{init})_{;l}(\3R_{init})^{;l}}
 {\3R_{init}^{2}}\right\}<\3R_{cr}(t, \tilde\phi_0),
\label{criterion2}
\end{equation}
where $\3R_{cr}(t, \tilde\phi_0)$ is the marginal value 
of the spatial curvature 
for inflation in the closed Friedmann model appears in Eq.(\ref{criterion02})  
and the parameters $\tilde\alpha$ and $\tilde\beta$ are 
\begin{equation}
	\tilde\alpha=(3\alpha-2), \quad 
	\tilde\beta=\frac{12\alpha-3\alpha\beta-7}{2(3\alpha-2)} . 
\end{equation}
The criterion (\ref{criterion2}) is a condition on the initial value of 
the spatial curvature and its spatial derivatives. 
Because $\tilde\alpha$ and $\tilde\beta$ are several numbers when 
$\tilde\phi_0=(5.0\sim 10.0)\kappa^{-1/2}$, 
the onset of inflation is occur 
if the initial value of spatial curvature $\3R_{init}$ is 
less than the critical value $\3R_{cr}(t, \tilde\phi_0)$
over a region which has several size of the local curvature
radius $(\3R_{init})^{-1/2}$.

Goldwirth and Piran studied how inhomogeneities influence the
inflationary epoch numerically in the case of spherical symmetry\cite{GP2}.
They conclude that the crucial feature necessary for inflation 
is a sufficiently high average value of the scalar field ( suitable 
value for inflation in homogeneous universe ) over a region of several 
Hubble radius. 
In order to compare their results with ours,
we impose $\Omega^2(x)$ to spherically symmetric form, 
$\Omega^2(\chi,\theta,\varphi)=\Omega^2(\chi)$, in the spherical 
coordinate where 
$h_{ij}=\mbox{diag}[1,\sin^2\chi,\sin^2\chi\sin^2\theta]$.
We consider $\Omega^2(\chi)$ in the form 
\begin{eqnarray}
\Omega^2(\chi) &=& \Omega_\pi^2
	+(\Omega_0^2-\Omega_\pi^2)\left[
         \frac{1-\exp\left[-\Delta^{-2}\cos^2\frac{\chi}{2}
 \right]}{1-\exp\left[-\Delta^{-2}\right]}\right] , 
 \label{gauss}
\end{eqnarray}
where $\Omega_0^2$ and $\Omega_\pi^2$ are the value of $\Omega$ at 
$\chi=0$ and $\chi=\pi$, respectively, 
and $\Delta$ is a parameter. 
We fix $\tilde{\phi}_0=5.0\kappa^{-1/2}$, for example, 
then $\Omega_{cr}^2\approx 0.70 m^{-2}$ (see Fig.\ref{fig:cs0}). 
The form of $\Omega^2$ for $(\Omega_0^2+ \Omega_\pi^2)/2= \Omega_{cr}^2$ 
and $\Omega_0^2 - \Omega_\pi^2= 4\times 10^{-7}m^{-2}$ 
is shown in Fig.\ref{fig:guss25}.
Varying $\Delta$, we show inflationary and recollapsing regions 
in the three-dimensional space (see Fig.\ref{fig:csgp25}).

At the origin $\chi=0$ universe can enter the inflationary phase 
whenever $\Delta$
is smaller than about 0.95. 
From Fig.\ref{fig:guss25}, we see that the region of a suitable value 
of $\Omega^2$ is 
$0\leq \chi\leq 1.85$. 
Comparing the physical length $L\sim a(t,\Omega)\chi$ for this region and 
the initial curvature radius $(\3R_{init})^{-1/2}\sim a(t,\Omega)/\sqrt{6}$, 
we get $L\sim 4.5 (\3R_{init})^{-1/2}$. 
In this case, inflation occurs at the origin when the initial value of 
spatial curvature is less than the critical value over the region 4.5 
times initial curvature radius.

\section{Conclusion}
\label{sec:con}

We consider a chaotic inflation model which is driven by 
a massive scalar field. 
In this model, we study how the initial inhomogeneities of the universe 
affect the onset of inflation by use of the gradient
expansion from a locally closed Friedmann universe.

At lowest order of our approximation, inhomogeneities of 
the metric and the scalar field are described by an arbitrary 
function $\Omega(x)$.
At next order, using the effective local scale factor $a_\eff(t,x)$, 
we investigate the effect of inhomogeneities of the universe 
for the onset of inflation. 

From numerical results, we obtain the condition for the onset
of inflation. 
The spatial curvature in the early stage should be less than the 
critical curvature value $\3R_{cr}(t, \tilde\phi_0)$ over a region which
has several times size of the local curvature radius 
$(\3R_{init}(t))^{-1/2}$ for the onset of inflation in the future.

We can compare the effect of inhomogeneities in the inflation model 
which is driven by a massive scalar field with the one in the model 
driven by a cosmological constant \cite{IIS}. 
The value of $\alpha$ in the scalar field model is larger than 
the one in the cosmological constant model. 
It means that the inflationary model driven by a massive 
scalar field is more sensitive to the inhomogeneities 
than one by a cosmological constant. 
%

\section*{Acknowledgements}
We would like to thank Professor A.Hosoya for continuous encouragement. 
This work was supported in part by the Monbusho Grant-in-Aid for
Scientific Research No. 5149.

\begin{appendix}

\section{Expression for the spatial curvature in the first order }
\label{sec:rij}

In the gradient expansion from the locally closed Friedmann universe, 
the spatial metric up to the second order of the spatial 
gradients is given in the form 
\begin{equation}
 \gamma_{ij}(t, x) = \0\gamma_{ij}(t, x)+\1\gamma_{ij}(t, x) , 
\end{equation}
where
\begin{eqnarray*}
 \0\gamma_{ij}(t, x) &=& a^2(t,\Omega) h_{ij}, \\
 \1\gamma_{ij}(t, x) &=& a^2(t,\Omega)\left[
 	\frac{1}{3}F(t,\Omega)\Oll h_{ij}
 	+\bar F(t,\Omega)\overline{\frac{\nabla_i\nabla^j\Omega}{\Omega}}
	\right.  \nonumber\\ 	
	&& \qquad\qquad +\frac{1}{3}G(t,\Omega) \OlOl h_{ij}
 	\left.+\bar G(t,\Omega)\overline{\frac{\nabla_i\Omega\nabla^j\Omega}{\Omega^2}}
	\right] . 
\end{eqnarray*}
The spatial Ricci curvature for the metric is 
\begin{eqnarray}
 \3\!\1R_i\!\!{}^j({\gamma_{kl}}) 
	= -\frac{1}{a^2}&&\left\{\frac{2}{3}\left[ 
  		 F-\bar F +2\frac{a'\Omega}{a}\right]
		\Oll \delta_i^{~j} \right. 
	+\frac{2}{3}\left[ G+\bar F -{\bar F}'\Omega
		 +{\bar F}^2 +\Omega^2\left(2\frac{a''}{a} 
		-\frac{a'^2}{a^2}\right)\right] \OlOl\delta_i^{~j} \nonumber \\
	&&+\left(\frac{a'\Omega}{a} \right)
	\overline{\frac{\nabla_i\nabla^j\Omega}{\Omega}} 
	+\left.\left[  2\bar G +2\bar F 
		-2{\bar F}'\Omega-\frac{5{\bar F}^2}{2}
 	-2\left(\frac{a'\Omega}{a}\right)^2 
	+\frac{a''\Omega^2}{a}\right]
 	\overline{\frac{\nabla_i\Omega\nabla^j\Omega}{\Omega^2}}\right\}. 
\label{eq:3Ricci}
\end{eqnarray}

\section{Equations for $\a', \a'', \phi_0', \phi_0''$ and $\overline{F}'$}
\label{sec:subeq}

We derive the equation for $a', a'', \phi_0', \phi_0''$
and $\overline{F}'$. 

From Eqs.(\ref{eq:sa0m}) and (\ref{eq:sa0c}) 
we obtain
\begin{eqnarray}
 \ddot{a}(t, \Omega) 
	&=& \frac{\dot{a}^2(t, \Omega)}{a(t, \Omega)} +\frac{1}{a(t, \Omega)}
	-\frac{\kappa}{2}a(t, \Omega)\dot{\phi}^2_0(t, \Omega). 
\label{eq:subeq0}
\end{eqnarray}

Differentiating Eqs.(\ref{eq:sphi0}) and (\ref{eq:subeq0})
with respect to $\Omega$,
we find the equations which $a'(t, \Omega)$ and $\phi'(t, \Omega)$ must satisfy 
\begin{eqnarray}
 \ddot{a}'&=& 2\frac{\dot{a}}{a}\dot{a}'-\left[
 \frac{\dot{a}^2+1}{a^2}+\frac{\kappa}{2}\dot{\phi}^2_0\right]a'
 -\kappa\dot{\phi}_0\dot{\phi}'_0 a, \label{eq:a'} \\
 \ddot{\phi}'_0 &=& -3\frac{\dot{a}}{a}\dot{\phi}'_0
 -\left(\frac{dV}{d\phi}\right)'\Big\vert_0 
 -3\frac{a\dot{a}'-\dot{a}a'}{a^2}\dot{\phi}_0.
 \label{eq:phi'}
\end{eqnarray}

Differentiating Eqs.(\ref{eq:a'}) and (\ref{eq:phi'}) again,
we find 
\begin{eqnarray}
 \ddot{a}''&=& 2\frac{\dot{a}}{a}\dot{a}''
 -\left[\frac{\dot{a}^2+1}{a^2}+\frac{\kappa}{2}\dot{\phi}^2_0
 \right]a''
+2\frac{\dot{a}'^2}{a}-4\frac{a'\dot{a}\dot{a}'}{a^2}
 +2\frac{a'^2(\dot{a}^2+1)}{a^3} \nonumber \\
 &&-2\kappa{a}'\dot{\phi}_0\dot{\phi}'_0
 -\kappa{a}\left(\dot{\phi}_0\dot{\phi}''_0+\dot{\phi}'^2_0\right),
  \label{eq:a''} \\
 \ddot{\phi}'' &=& -3\frac{\dot{a}}{a}\dot{\phi}''_0
 -\left(\frac{dV}{d\phi}\right)''\Big\vert_0 
 -6\dot{\phi}'_0\frac{a\dot{a}'-\dot{a}a'}{a^2}
 -3\dot{\phi}_0\left[\frac{\dot{a}''}{a}-2\frac{a'\dot{a}'}{a^2}
	-\frac{a''\dot{a}}{a^2}+2\frac{\dot{a}a'^2}{a^3}\right].
 \label{eq:phi''}
\end{eqnarray}

Similarly from Eq.(\ref{eq:smc1}),
we find that $\overline{F}'$ must satisfy 
\begin{eqnarray}
 \dot{\overline{F}}'&=& 2\frac{a\dot{a}'-a'\dot{a}}{a^2}
 +2\Omega\left[\frac{\dot{a}''}{a}-2\frac{a'\dot{a}'}{a^2}
	-\frac{a''\dot{a}}{a^2}+2\frac{\dot{a}a'^2}{a^3}\right]
 \nonumber \\
 &&+\kappa\left(\dot{\phi}_0\phi'_0+\dot{\phi}'_0\phi'_0\Omega
 +\dot{\phi}_0\phi''_0\Omega\right). \label{eq:overF'}
\end{eqnarray}

\section{The spatial tensors}
\label{sec:tensor}

The spatial tensors $(\nabla^{(2p)}\Omega)^A_{ij}$ 
which contain $2p$ spatial gradients of $\Omega$ 
are given as follows.

For $(\nabla^{(2)}\Omega)^A_{ij}$
\begin{eqnarray*}
 \Oij \quad \mbox{and} \quad \OiOj . 
\end{eqnarray*}
For $(\nabla^{(4)}\Omega)^A_{ij}$  
\begin{eqnarray*}
 \frac{\nabla_i\nabla_j\nabla_l\nabla^l\Omega}{\Omega}, \\
 \frac{\nabla_i\nabla_j\Omega\nabla_l\nabla^l\Omega}{\Omega^2},\quad
 \frac{\nabla_i\Omega\nabla_j\nabla_l\nabla^l\Omega}{\Omega^2}, \quad
 \frac{\nabla_l\nabla_i\nabla_j\Omega\nabla^l\Omega}{\Omega^2}, \quad
 \frac{\nabla_i\nabla_j(\nabla_l\Omega\nabla^l\Omega)}{\Omega^2}, \\
 \frac{\nabla_i\Omega\nabla_j\Omega\nabla_l\nabla^l\Omega}{\Omega^3},\quad
 \frac{\nabla_i\nabla_j\Omega\nabla_l\Omega\nabla^l\Omega}{\Omega^3},\quad
 \frac{\nabla_i\Omega\nabla_j(\nabla_l\Omega\nabla^l\Omega)}{\Omega^3},\\
 \frac{\nabla_i\Omega\nabla_j\Omega\nabla_l\Omega\nabla^l\Omega}{\Omega^4}. 
\end{eqnarray*}

\end{appendix}



\vskip 1.0cm
\begin{center}
{\bf Figure Captions}
\end{center}
\vskip 0.25cm

\begin{figure} 
\begin{center}
\figurenum{1}
  \leavevmode
  \epsfbox{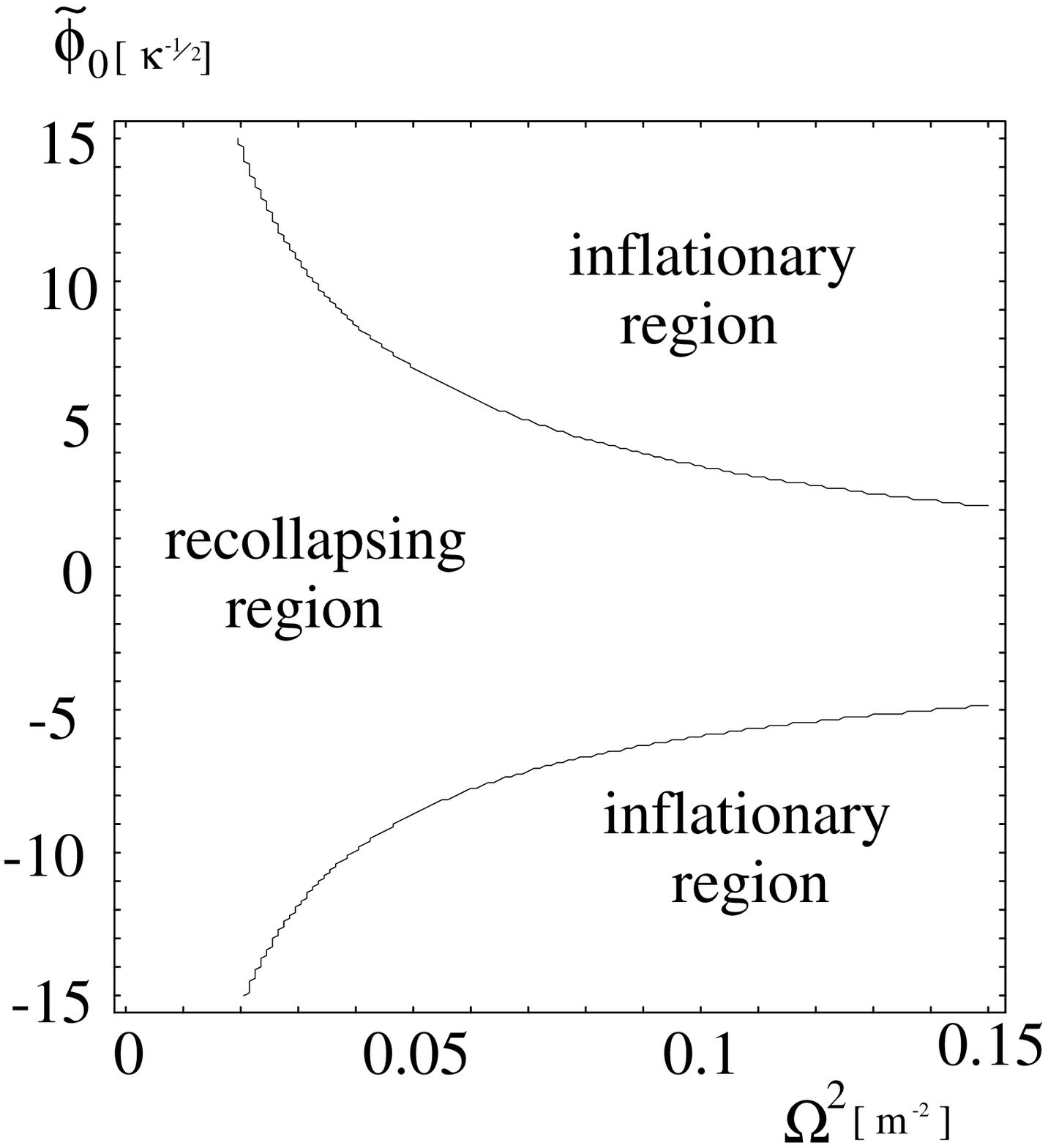}
\caption{}
\label{fig:cs0}
\end{center}
\end{figure}

\noindent
Fig.\ref{fig:cs0}. \\
\noindent
In the lowest order approximation, an inflationary region 
and recollapsing region are shown in $(\Omega^2, \tilde{\phi}_0)$ plane.

\vskip 0.5cm

\begin{figure}
\begin{center}
\figurenum{2-(i)}
  \leavevmode
  \epsfbox{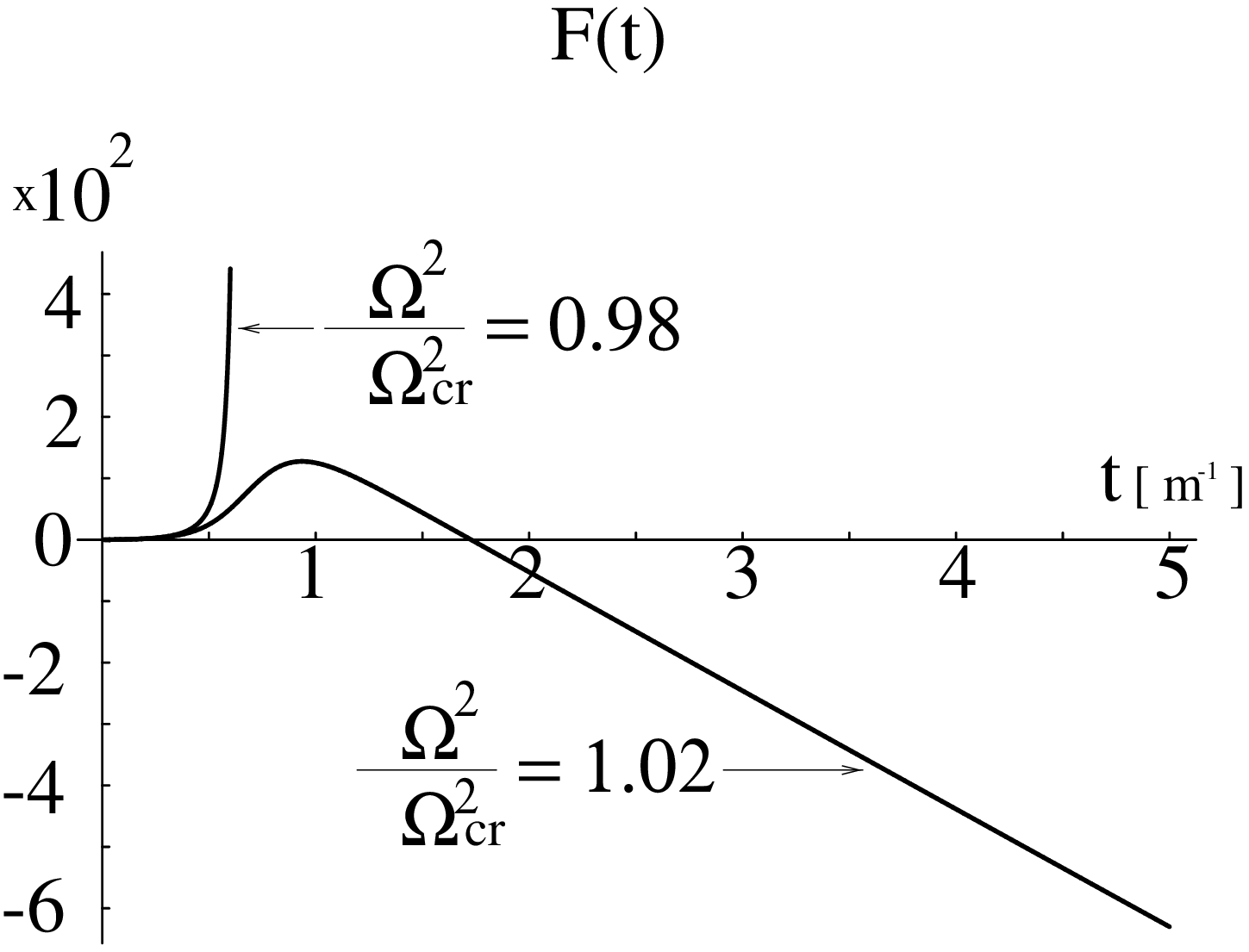}
\caption{}
\label{fig:F}
\end{center}
\end{figure}

\begin{figure}
\begin{center}
\figurenum{2-(ii)}
  \leavevmode
  \epsfbox{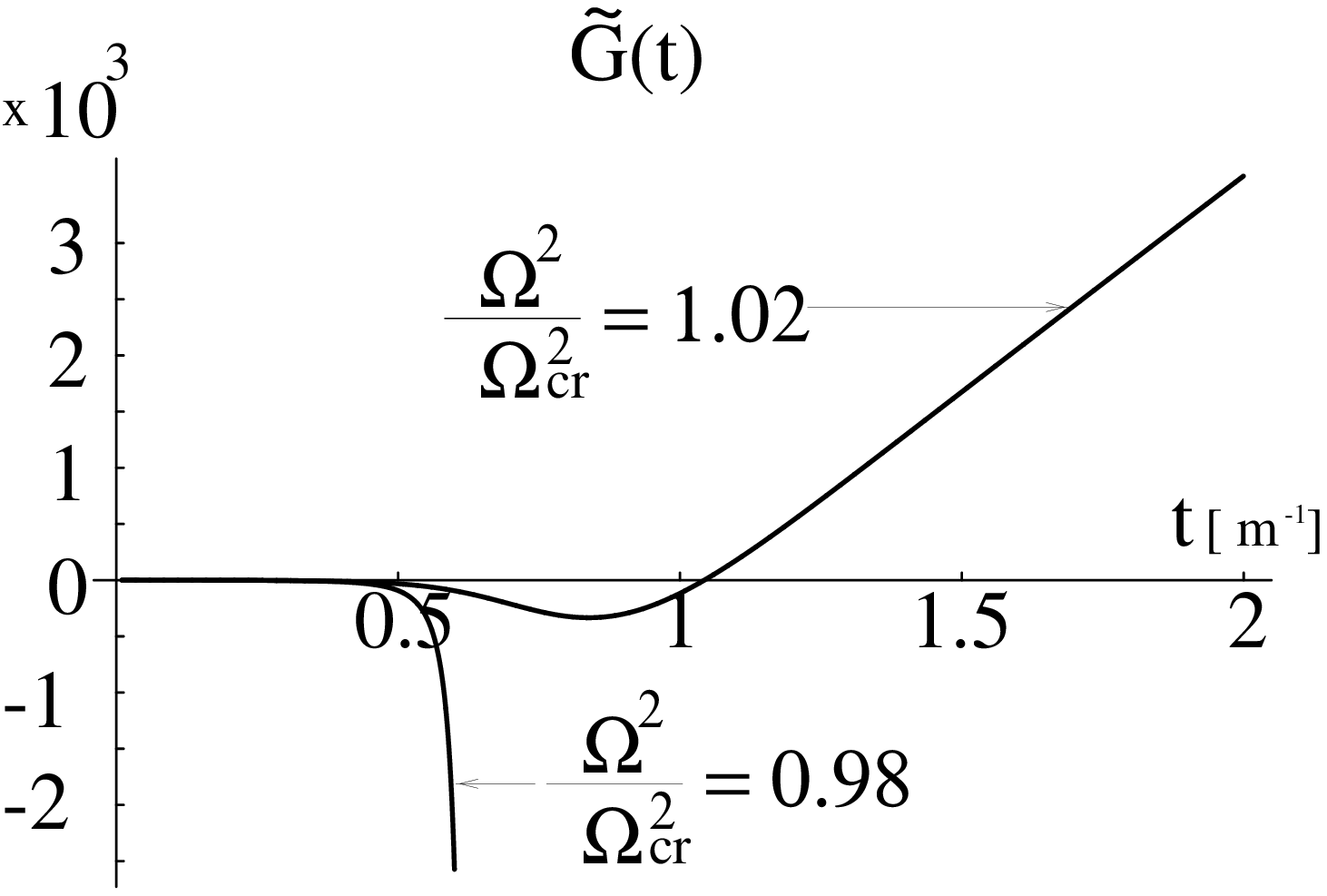}
\caption{}
\label{fig:G}
\end{center}
\end{figure}

\begin{figure}
\begin{center}
\figurenum{2-(iii)}
  \leavevmode
  \epsfbox{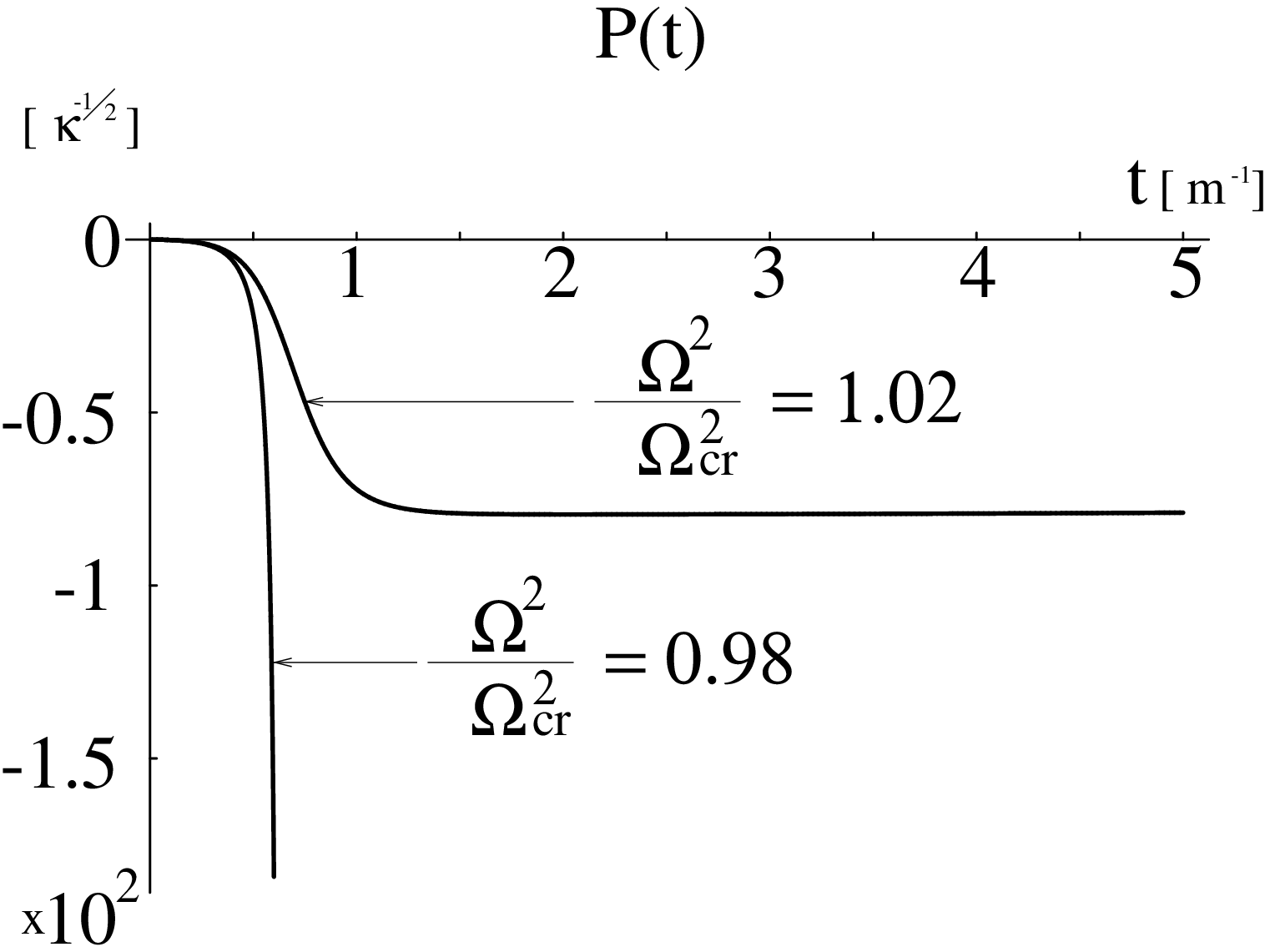}
\caption{}
\label{fig:P}
\end{center}
\end{figure}

\begin{figure}
\begin{center}
\figurenum{2-(iv)}
  \leavevmode
  \epsfbox{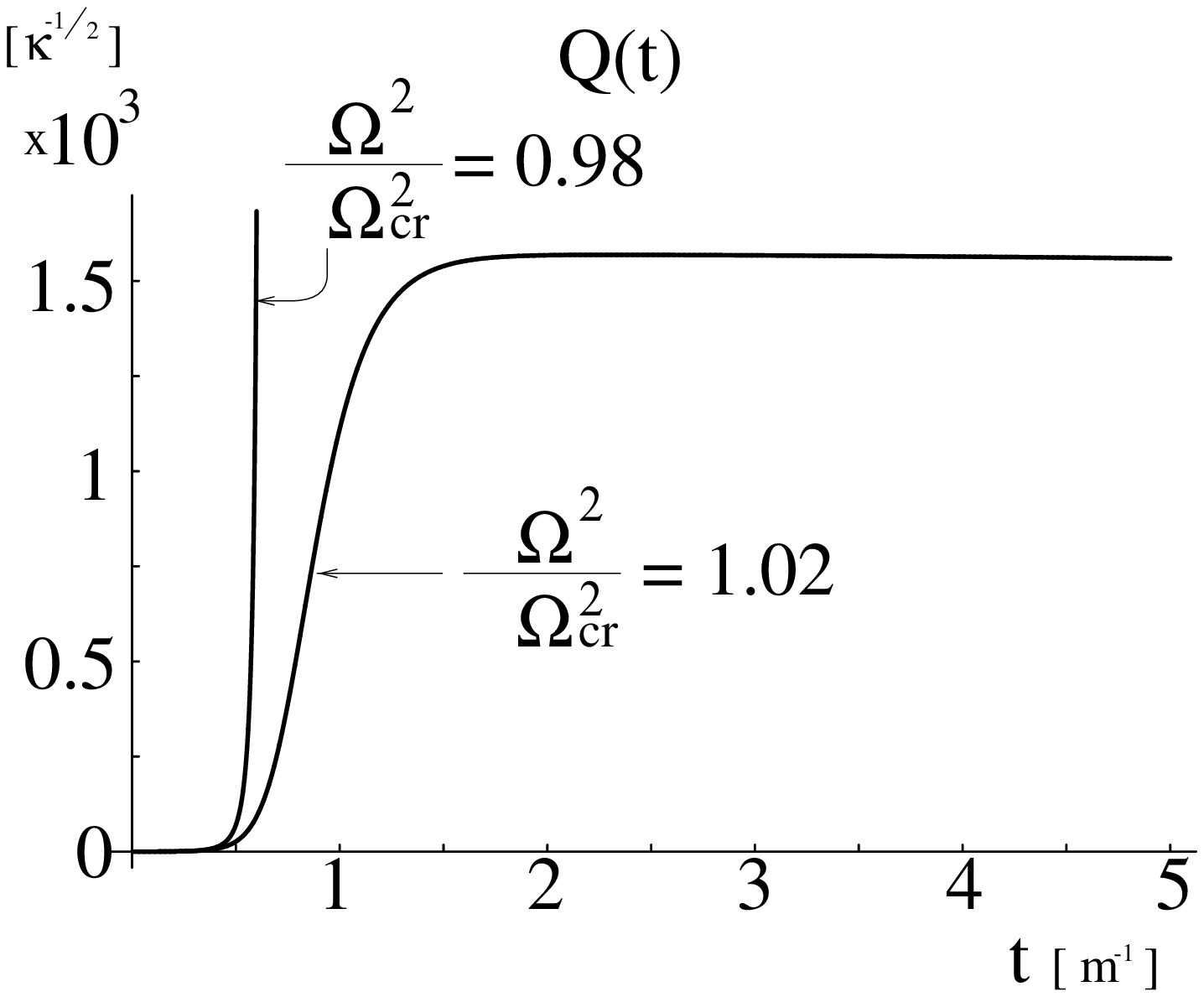}
\caption{}
\label{fig:Q}
\end{center}
\end{figure}

\noindent
Figs.\ref{fig:F}--\ref{fig:Q}. \\
\noindent
Typical examples of the time evolution of 
(i)$F(t, \Omega)$, (ii)$\tilde{G}(t, \Omega)$, (iii)$P(t, \Omega)$ 
and (iv)$Q(t, \Omega)$.

\begin{figure}
\begin{center}
\figurenum{3-(i)}
  \leavevmode
  \epsfbox{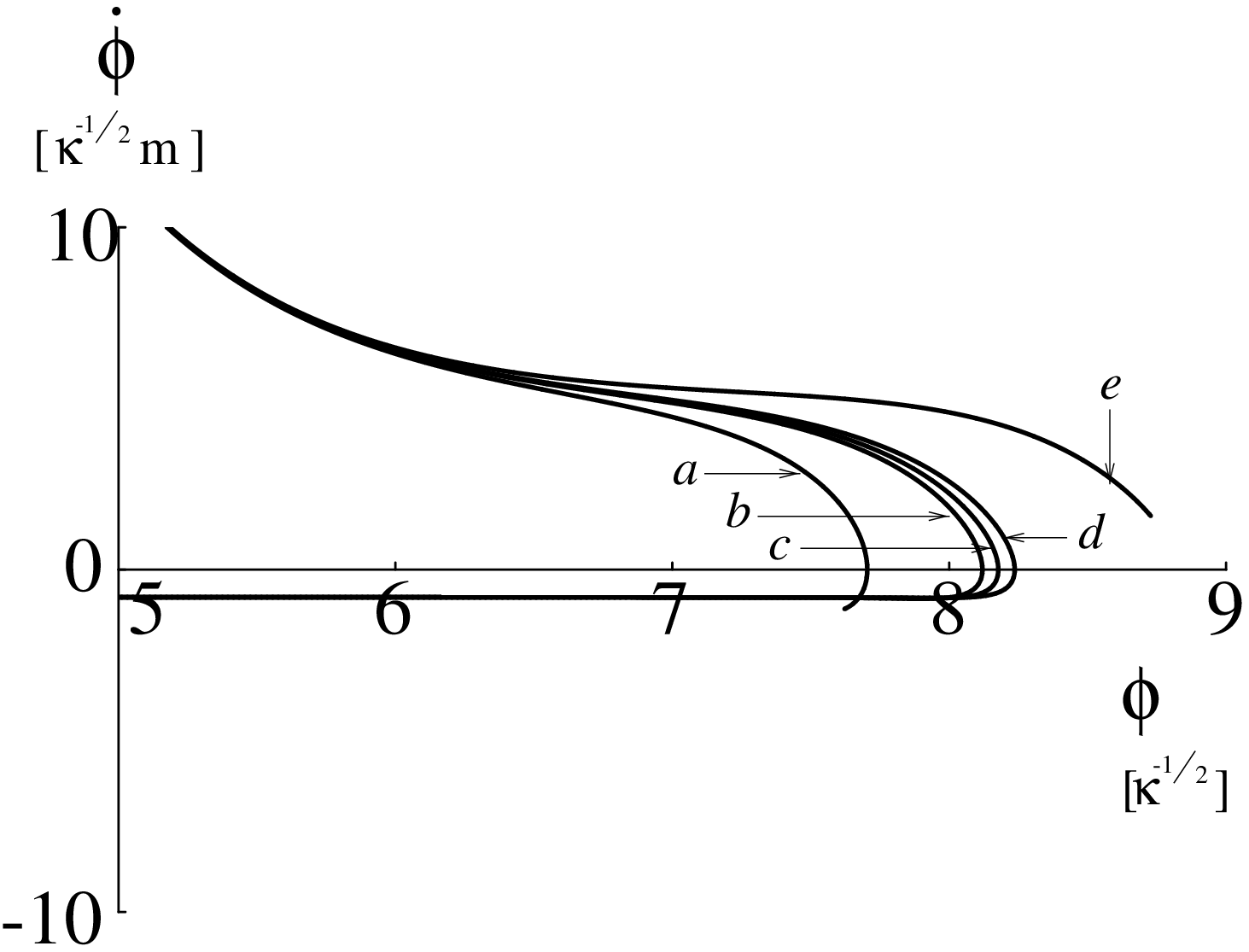}
\caption{}
\label{fig:phia}
\end{center}
\end{figure}

\begin{figure}
\begin{center}
\figurenum{3-(ii)}
  \leavevmode
  \epsfbox{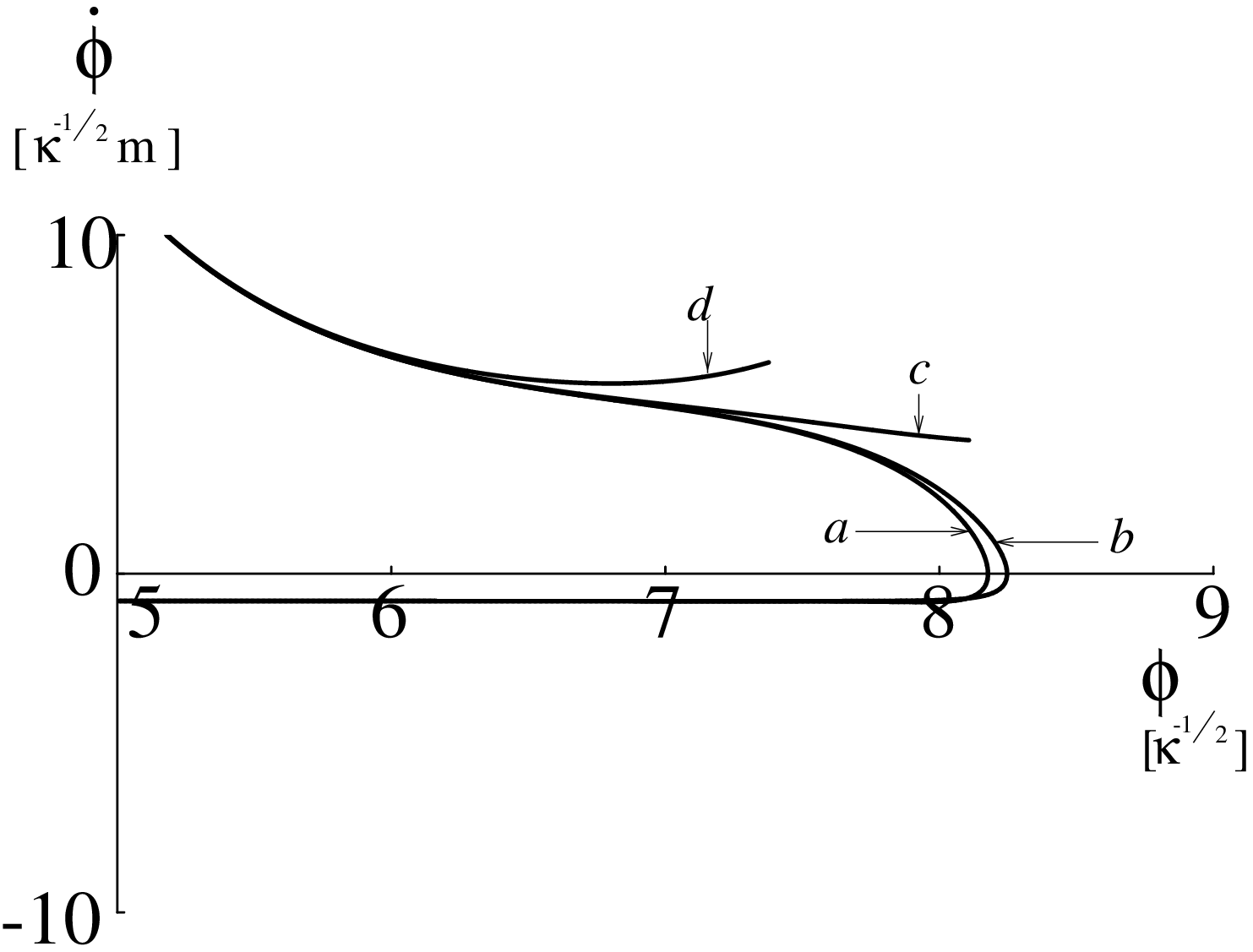}
\caption{}
\label{fig:phib}
\end{center}
\end{figure}

\noindent
Figs.\ref{fig:phia}--\ref{fig:phib}. \\
\noindent
The trajectories in $(\phi,\dot\phi)$ space. 

(i) $\nabla_{l}\Omega\nabla^{l}\Omega/\Omega^2=0$ and
$\nabla_{l}\nabla^{l}\Omega/\Omega=$ 0.01(a), 0.001(b), 0(c),
-0.001(d) and -0.01(e). 

(ii) $\nabla_{l}\nabla^{l}\Omega/\Omega=0$ and
$\nabla_{l}\Omega\nabla^{l}\Omega/\Omega^2=$ 0(a), 0.0001(b),
0.001(c) and 0.01(d).

At the endpoint of the trajectory the approximation breaks down. 
\vskip 0.5cm

\begin{figure}
\begin{center}
\figurenum{4-(i)}
  \leavevmode
  \epsfbox{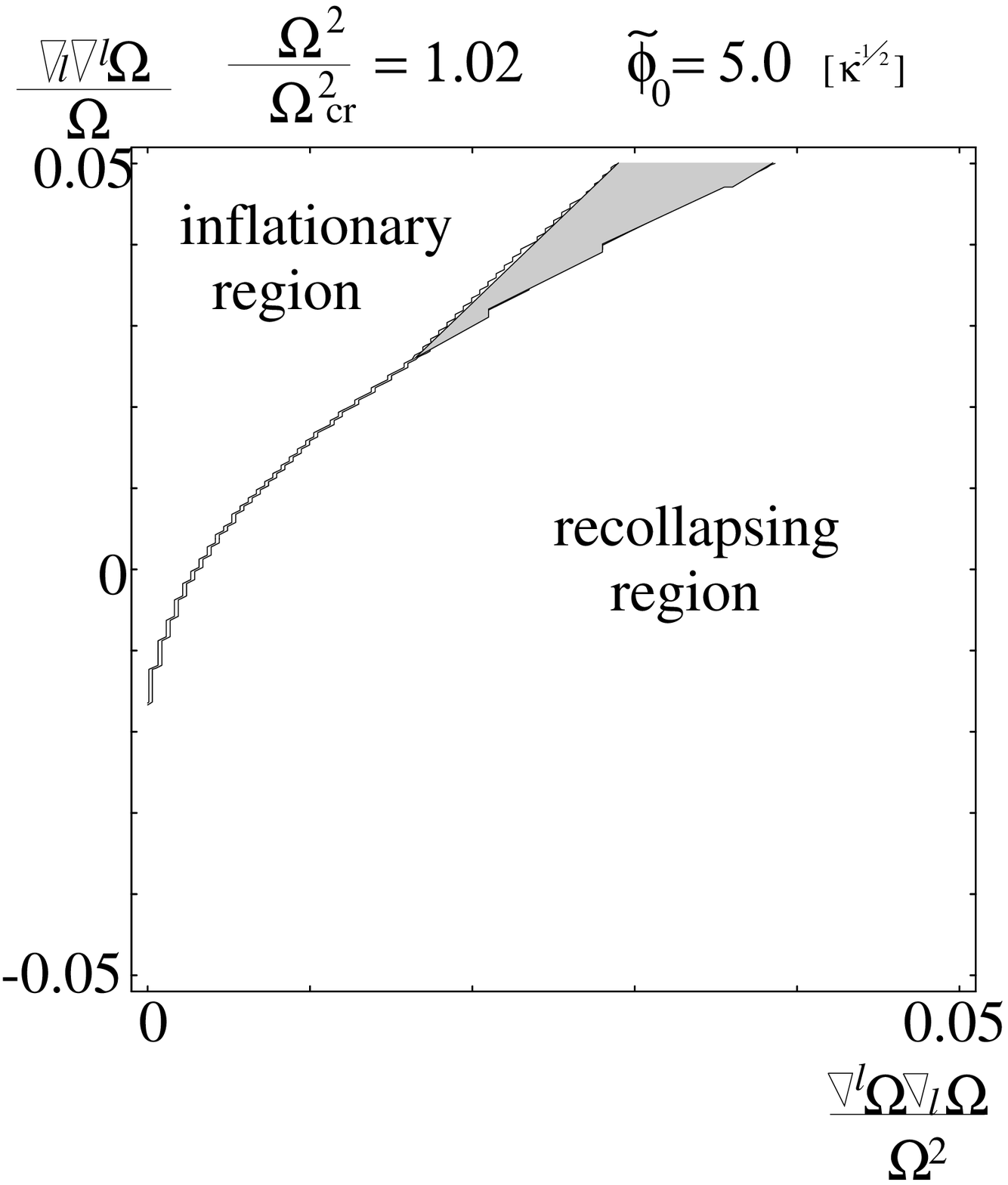}
\caption{}
\label{fig:cs2502}
\end{center}
\end{figure}

\begin{figure}
\begin{center}
\figurenum{4-(ii)}
  \leavevmode
  \epsfbox{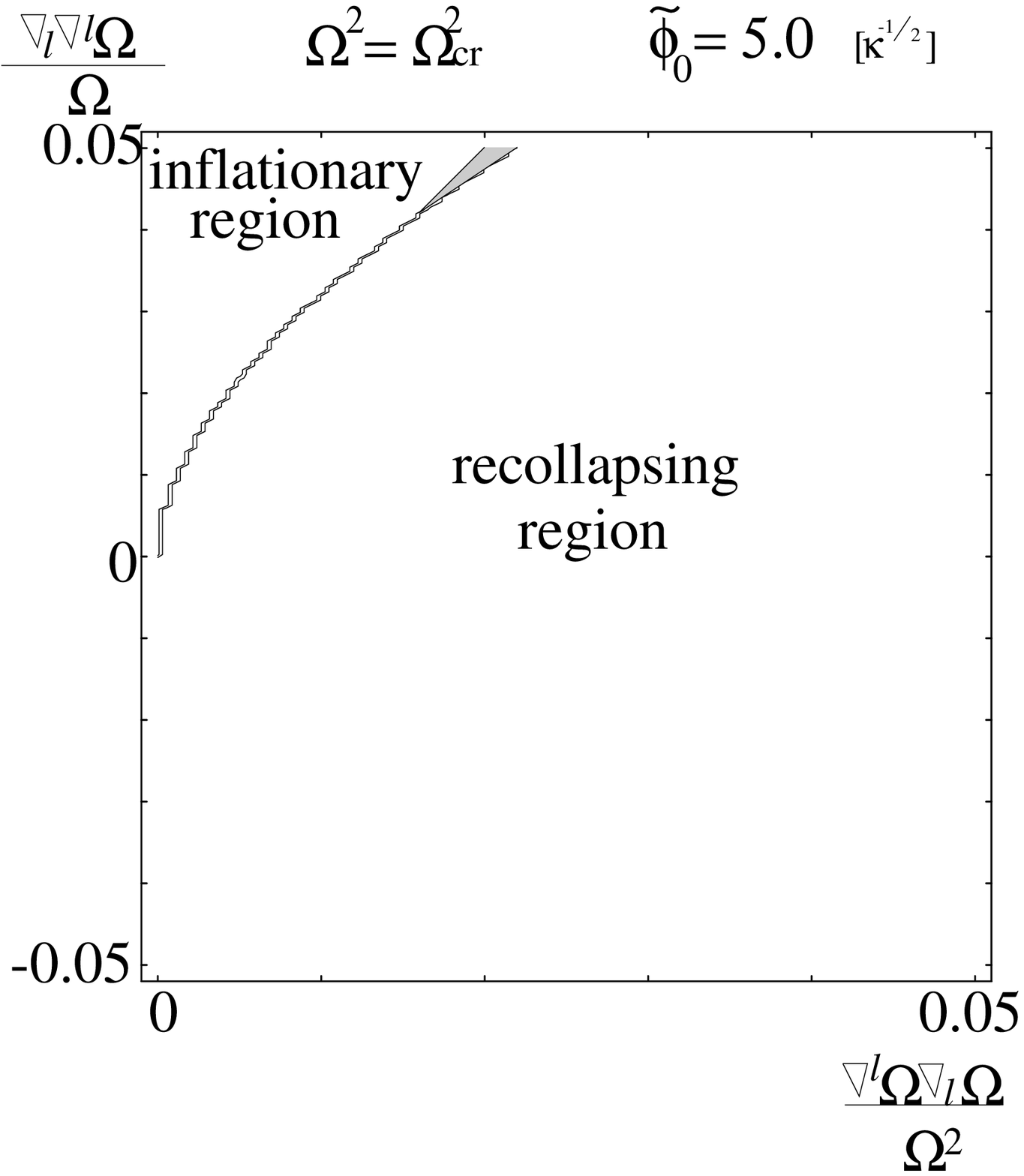}
\caption{}
\label{fig:cs25}
\end{center}
\end{figure}

\begin{figure}
\begin{center}
\figurenum{4-(iii)}
  \leavevmode
  \epsfbox{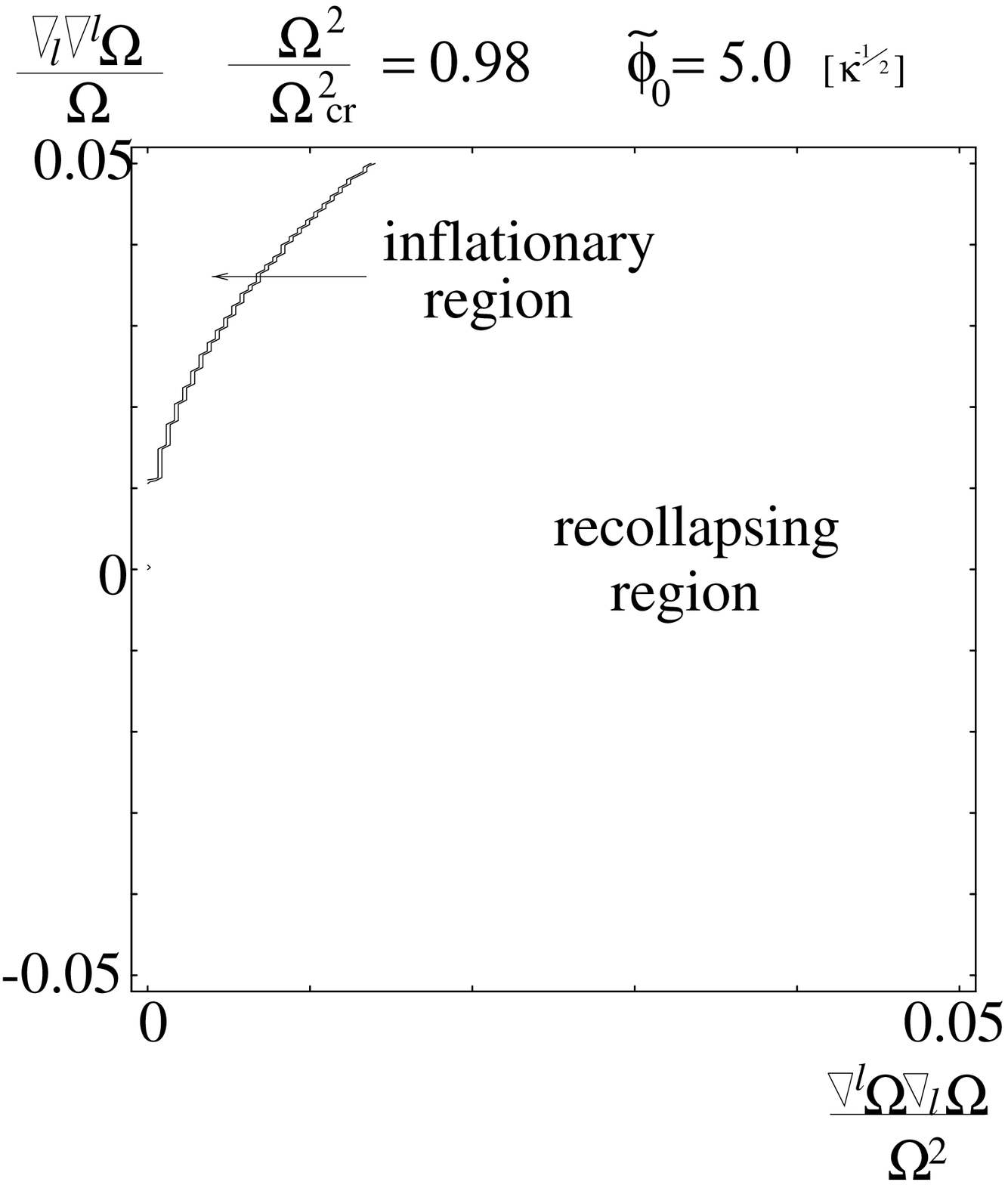}
\caption{}
\label{fig:cs2598}
\end{center}
\end{figure}

\noindent
Figs.\ref{fig:cs2502}--\ref{fig:cs2598} \\
\noindent
An inflationary region and a recollapsing region 
for fixed $\tilde\phi_0=5.0\kappa^{-1/2}$ ($\Omega^2_{cr}\approx 0.70m^{-2}$)
are shown 
in the parameter space ( $\nabla_l\nabla^l\Omega/\Omega$, 
$\nabla_l\Omega\nabla^l\Omega/\Omega^2$ ). 
A meshed region is a region where the approximation breaks down.
\vskip 0.5cm

\begin{figure}
\begin{center}
\figurenum{5}
  \leavevmode
  \epsfbox{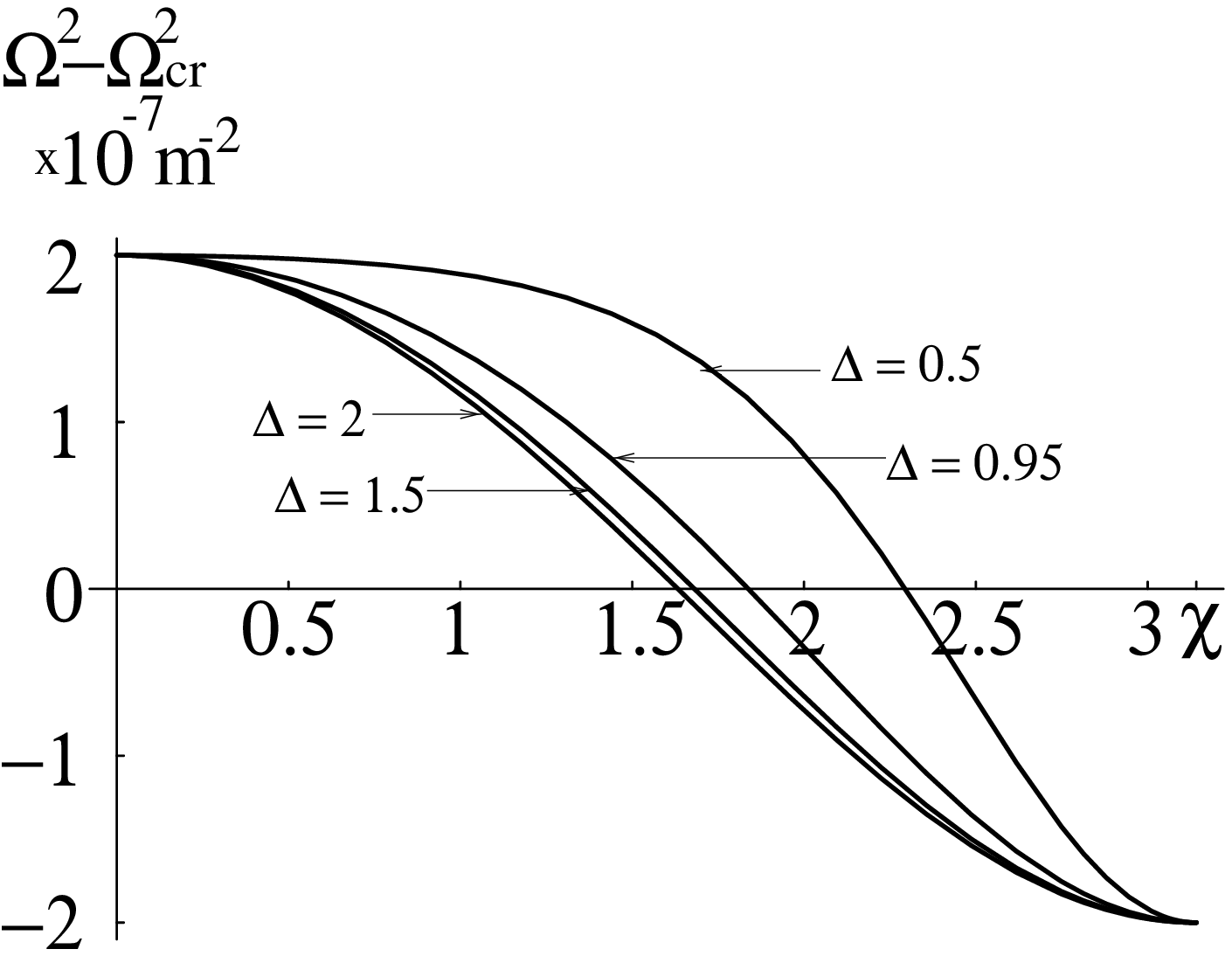}
\caption{}
\label{fig:guss25}
\end{center}
\end{figure}

\noindent
Fig.\ref{fig:guss25}. \\
\noindent
The shape of $\Omega^2(\chi)$ for varying $\Delta$. 

\vskip 0.5cm

\begin{figure}
\begin{center}
\figurenum{6}
  \leavevmode
  \epsfbox{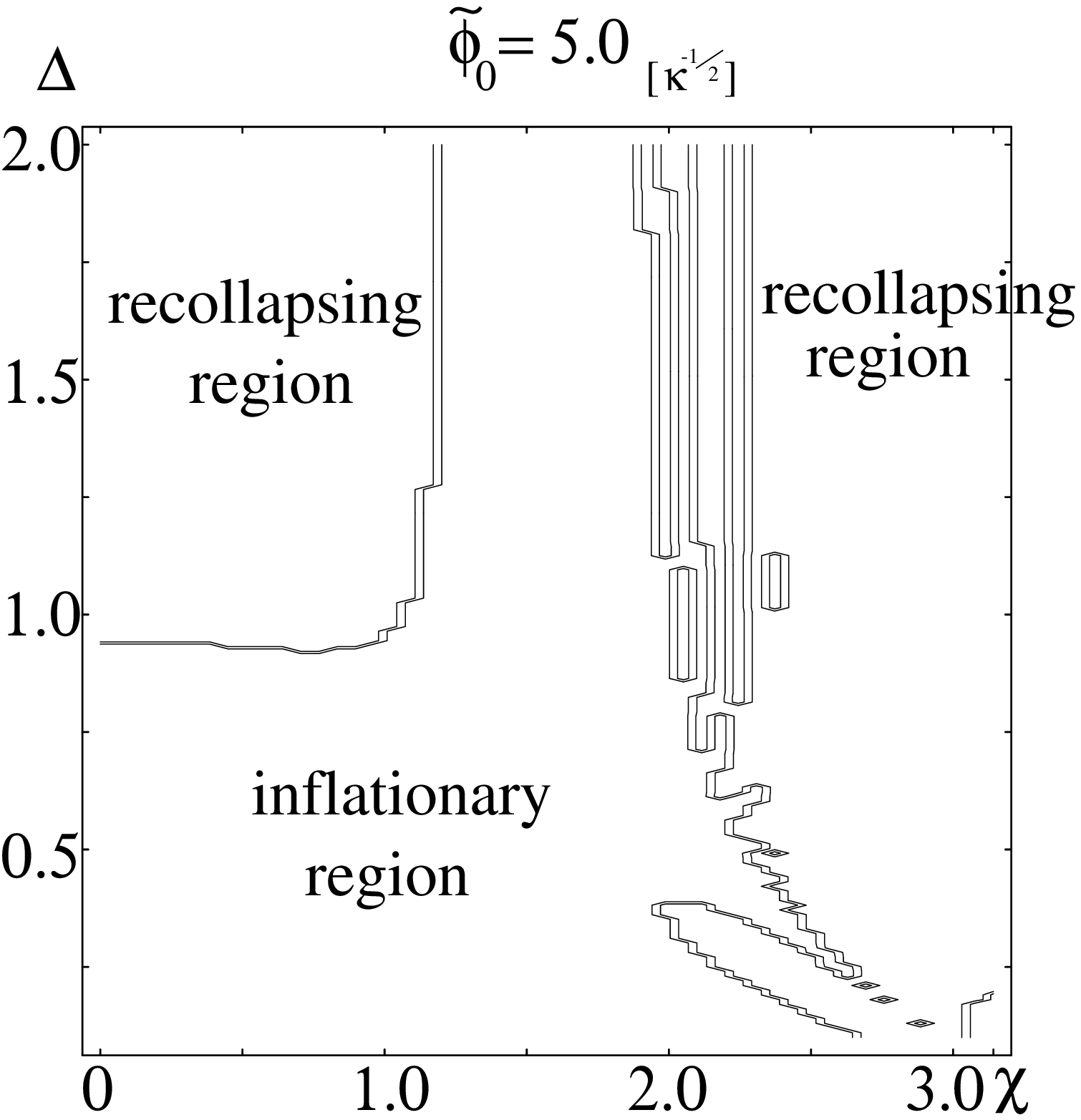}
\caption{}
\label{fig:csgp25}
\end{center}
\end{figure}

\noindent
Fig.\ref{fig:csgp25}. \\
\noindent
An inflationary region and a recollapsing region 
in the three-dimensional space 
for fixed $\tilde\phi_0=5.0\kappa^{-1/2}$ are shown
in the case that $\Omega^2$ is given in Fig.\ref{fig:guss25}. 

\end{document}